# WEAKLY SUPERVISED CRNN SYSTEM FOR SOUND EVENT DETECTION WITH LARGE-SCALE UNLABELED IN-DOMAIN DATA

*Dezhi Wang[1*], Lilun Zhang[1], Changchun Bao[1], Kele Xu[2], Boqing Zhu[2], Qiuqiang Kong[3]*

[1]College of Meteorology and Oceanography, National University of Defense Technology, Changsha, 410073, China
[2]School of Computer, National University of Defense Technology, Changsha, 410073, China
[3]CVSSP, University of Surrey, UK

## ABSTRACT

Sound event detection (SED) is typically posed as a supervised learning problem requiring training data with strong temporal labels of sound events. However, the production of datasets with strong labels normally requires unaffordable labor cost. It limits the practical application of supervised SED methods. The recent advances in SED approaches focuses on detecting sound events by taking advantages of weakly labeled or unlabeled training data. In this paper, we propose a joint framework to solve the SED task using large-scale unlabeled in-domain data. In particular, a state-of-the-art general audio tagging model is first employed to predict weak labels for unlabeled data. On the other hand, a weakly supervised architecture based on the convolutional recurrent neural network (CRNN) is developed to solve the strong annotations of sound events with the aid of the unlabeled data with predicted labels. It is found that the SED performance generally increases as more unlabeled data is added into the training. To address the noisy label problem of unlabeled data, an ensemble strategy is applied to increase the system robustness. The proposed system is evaluated on the SED dataset of DCASE 2018 challenge. It reaches a F1-score of 21.0%, resulting in an improvement of 10% over the baseline system.

## 1. INTRODUCTION

Great attention has been paid to developing advanced approaches to understand the sounds of everyday life in the contexts of practical applications such as smart cars[1], surveillance [2, 3], healthcare [4]. Sound event detection (SED) has been studied to automatically achieve the strong temporal annotations for the occurrences of sound events in an audio recording. As the rapid development in recent years, deep learning methods have become the main approaches to solve the SED task, especially in the IEEE audio event and scene detection challenges (DCASE)[1, 5].

Ideally strongly labeled audio data is preferable for SED model development in terms of supervised learning manner [6, 7]. However, the cost of producing strongly labeled dataset is remarkably high for the data-intensive deep-learning methods. Therefore, it is desirable to develop weakly supervised SED methods that can exploit weakly labeled data and unlabeled data effectively. As the release of Google AudioSet [8] that consists of approximately 2-million weakly labeled short audio clips, the progress of weakly supervised learning SED approaches has been significantly accelerated.

**1.1 Sound event detection by weakly supervised learning**

Apart from supervised learning, the weakly supervised SED approaches can be normally grouped into three domains:

(1) The methods developed under inexact supervision [9] to predict strong annotations of sound events using weakly labeled training data. In terms of using the weak labels for SED, there are mainly four ways. One way is to directly assign the clip-level labels (weak labels) to all the frame-level segments (strong labels) to train the model [10, 11], which may introduce noise to the frame-level labels. Another way is to do the source separation at first and obtain the frame-level labels [12, 13] based on the separated sources. The third way is to use the attention mechanisms [14, 15] to learn the relationship between frame-level labels and clips-level labels in training process. The fourth way is to formulate the weak-label SED job as a multiple instance learning (MIL) problem [16-18], where the audio clips are treated as bags and audio segments are treated as instances.

(2) The methods implemented as incomplete supervision [9] to make use of unlabeled training data to increase performance. The semi-supervised learning is the major technique for this purpose [19-21]. Moreover, virtual adversarial training (VAT) can also be used to process un-labelled data [20, 22]. Mean teacher algorithm can improve the performance of semi-supervised SED system by using unlabeled data [23]. The application of ensemble mechanisms is a good strategy to make the semi-supervised learning more reliable [24].

This work was funded by the National Natural Science Foundation of China (No. 61806214) and Scientific Research Project of NUDT (No.ZK17-03-31).

(3) The methods used to deal with the noisy labels of training data in an inaccurate supervision [9] manner. To address the noisy label problem in SED, a practical idea is to identify the mislabeled samples and make some corrections. The majority voting strategy and sample re-weight strategy [25] employed in the ensemble methods are widely used techniques. Some effort is also made to implement an iteratively fine-tuning framework by means of self-verifying the training observations [26].

## 1.2 Our contributions

In this paper, we aim to develop a scalable system based on a CRNN framework using the well-developed neural networks for weakly-supervised SED. The main contributions of our work can be summarized as: (1) In order to make use of unlabeled training data, we use our audio tagging system (NUDT system) ranked as the top in the DCASE 2018 challenge [27] to contribute on predicting the weak labels for unlabeled data more effectively. (2) We explore to integrate well-developed CNN architectures such as ResNet [28] and Xception [29] into the CRNN framework for more effective feature extraction. The SED performance significantly benefits from fine-tuning these pre-trained CNN models. (3) To address influence of using unlabeled training data on SED results, we make a comparative study by respectively adding unlabeled data with different confidence levels into the training. (4) In order to tackle the noisy label problem caused by using unlabeled training data, we apply a model ensemble technique to increase the robustness of proposed system. Our code can be referred to https://github.com/Blank-Wang/DCASE2018-Task4.

## 2. METHODOLOGY

### 2.1 Audio tagging for unlabeled in-domain training data

In order to explore the possibility of making use of a large amount of unlabeled training data, we utilize our proposed system [27] (called NUDT system) for the general audio tagging task in DCASE 2018 to predict the weak labels for unlabeled in-domain data (as shown in Figure 1). The system[1] applies fine-tuning on several popular pre-trained CNN architectures and utilizes the ensemble learning to enhance the performance. It is reported to have achieved the state-of-the-art performance in the DCASE 2018 audio tagging task. In the label prediction process, a 5-fold cross-validation is employed on the weakly labeled training dataset to fine tune the system. Since SED is a multi-label learning task, 3 threshold values are applied on the system softmax outputs (range in [0, 1]) to keep up to 3 classes of sound events in the predicted weak label for each unlabeled audio clip. The 3 thresholds are corresponding to 3 types of weak labels, i.e. 1-type (an audio clip label only has 1 class of sound event), 2-type (a clip label has 2 classes of sound events) and 3-type (a clip label has 3 classes of sound events). This setting is reasonable since there is few clip has more than 3 classes of sound events in the dataset [8]. After the unlabeled in-domain data are labeled by the audio tagging system, the distributions of weak labels under different thresholds can be illustrated as shown in Table 1. It is shown that the number of weakly labeled training data is actually significantly extended.

### 2.2 Weakly-supervised SED by a CRNN architecture

As shown in Figure 1, after the weak labels are obtained for the unlabeled in-domain training data, these data can be added together with the original weakly labeled data for the training of SED system. In this paper, a CRNN-based SED system is developed to predict the strong annotations of sound events in the audio clips. The well-developed ResNet50 or Xception model (without the top layers) is directly integrated into the CRNN framework as the CNN component to more effectively extract features from the time-frequency representations of input data. The ResNet or Xception models can take advantages of pre-trained model weights on 'ImageNet' and only need to be fine-tuned on a relatively small training dataset.

It should be noted that the ResNet or Xception model has been slightly modified by reducing the stride parameters in the time axis for the pooling and convolutional layers. In this way, the time-step information is sufficiently kept for a SED required time resolution. Following the CNN component, a reshape layer and a 1-D convolutional layer are applied to change the shape of CNN features before input to the RNN blocks. A gated bi-directional GRU layer is used as the RNN component, where GRU outputs are connected to a gated unit using 'sigmoid' and 'tanh' as activation functions. This kind of grated unit is the same one applied in Google WaveNet, which is believed to be more appropriate for audio signal processing [30]. Following the gated GRU layer, an additional feed-forward neural network is used to introduce the attention mechanism [14]. Finally, the SED strong annotations are obtained from the dense layer with 'sigmoid' activation.

### 2.3 Data augmentation based on mixup

A mixup technique [31] for data augmentation is applied in training process, which is believed to have some benefits on the model generalization to reduce over-fitting. It should be noted that both the raw wave signals and their time-frequency representations can be used as the mixup samples. In our experiment, we apply the mixup with alpha value of 0.2 [31] on the samples of time-frequency features.

### 2.4 Class balance

The class imbalance problem can lead the model to pay more attention to the classes with more training samples and neglect to learn from the classes with less samples. This problem is relatively significant for DCASE 2018 dataset [5]. We utilize a way to do the class balancing by limiting that the frequency of samples from the major classes to be randomly selected in a batch is at most 6 times than that of samples from the minor classes [14]. We also try another way to solve the class imbalance problem by re-weighing the class weights in the cost function during training, which can be conveniently implemented in Keras.

---
[1] The code in https://github.com/Cocoxili/DCASE2018Task2/

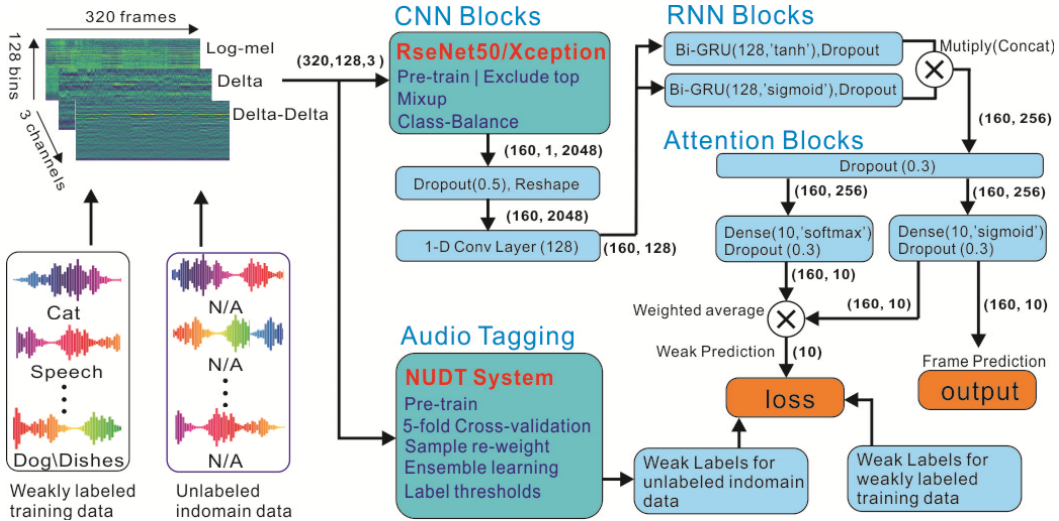

**Fig. 1.** The architecture of the overall system

### 2.5 Threshold adjustment

A threshold is needed in order to decide the existence of a sound event in an audio clip. It is important to choose a specific threshold for each class in the given task for SED. We manually adjust the threshold for each class on the testing dataset to obtain the optimal values, which are used to achieve SED results on evaluation dataset.

### 2.6 Model ensemble

Model ensemble can reduce the effect from the noise of weak labels predicted by the audio tagging system, which is an important technique to improve the system robustness and accuracy. In this work, we utilize the weighted average strategy to ensemble the outputs of our SED systems with different configurations.

## 3. EXPERIMENTS

### 3.1 Datasets and pre-processing

The dataset used is from 2018 DCASE challenge on large-scale weakly labeled sound event detection in domestic environments [8], which contains a training dataset (including 1578 weakly labeled clips, 14412 unlabeled in-domain clips and 39999 unlabeled out-of-domain clips), a testing dataset of 288 clips with strong labels and an evaluation dataset of 880 clips. We first down-sample all these audio clips from 44.1 kHz to 22.05 kHz and transform the wave forms into log-mel energies with 128 filter banks, a 2048 frame window and a frame shift of 684. Then the 1st and 2nd-order delta features of the log-mel features are produced and all these 3 parts of features are stacked together as a 3-channel feature representation used as the input of system. It should be noted that the out-of-domain unlabeled data is not used in this work. We argue that the usage of out-of-domain data may introduce extra noise especially when its distribution is significantly different.

**Table 1.** Distribution of weak labels in training datasets under different threshold values

| Dataset | Total | 1-type | 2-type | 3-type | 4-type | none |
|---|---|---|---|---|---|---|
| (a) wt | **1546** | 965 | 508 | 73 | 0 | 0 |
| (b) wt-0.99-0.47-0.28 | **8881** | 8219 | 579 | 83 | 0 | 0 |
| (c) wt-0.95-0.45-0.26 | **9259** | 8538 | 628 | 93 | 0 | 0 |
| (d) wt-0.9-0.43-0.23 | **9549** | 8702 | 717 | 130 | 0 | 0 |
| (e) wt-0.85-0.42-0.22 | **9810** | 8892 | 766 | 152 | 0 | 0 |
| (f) wt-0.8-0.4-0.2 | **11688** | 9804 | 1373 | 511 | 0 | 0 |
| test | **288** | 173 | 98 | 17 | 0 | 0 |
| evaluation | **880** | 420 | 362 | 50 | 2 | 46 |

e.g. wt-0.99-0.47-0.28' where 'wt' indicates the original weakly labeled training data, '0.99','0.47'and'0.28' respectively indicates the thresholds used to select the weak labels that only have 1-class, 2-class and 3-class sound events for unlabeled in-domain data.

### 3.2 Experimental setup

Different configurations of the proposed architecture are tested in this work. The audio tagging process is carried out based on a 5-fold cross-validation setup and the performance is validated on the test dataset in order to obtain the best model. In the SED process, both ResNet50 and Xception models with pre-trained weights on the 'ImageNet' are respectively used as the CNN component in the CRNN framework. The 'adam' algorithm is applied as the optimizer using a learning rate of 0.001 with the aid of an early-stopping technique using a patience of 7 and a maximum number of epochs that is 50. A batch size of 20 is used based on the hardware capacity. The model is trained on Keras accelerated by GTX 1080Ti GPU card. The SED performance is evaluated by means of class-wise F1 measure, which is calculated by the official sed_val package [1] with a 200ms collar on onsets and a 200ms/20% collar on offsets.

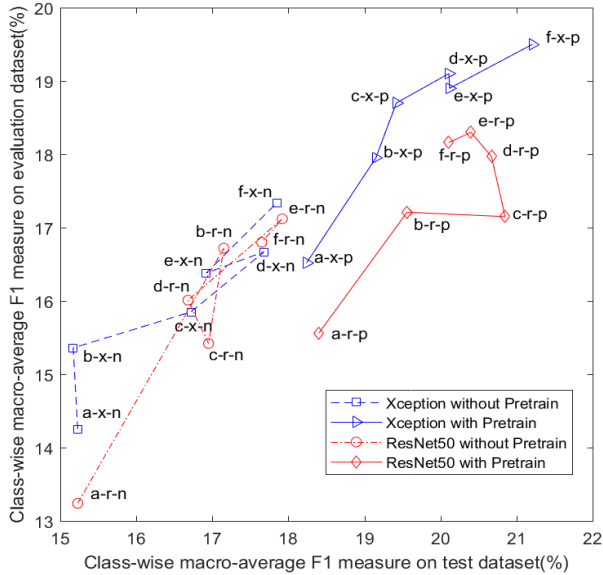

**Fig. 2.** SED F1 results on different training datasets ('a ~ f'). Notation 'r' and 'x' respectively indicate the ResNet50-based and Xception-based SED systems. 'n' indicates there is no pre-train applied while 'p' indicates using the pre-train

## 4. RESULTS

The proposed SED system is respectively trained on a series of training datasets as shown in Table 1. As illustrated in Figure 2, the results of the SED system under different combinations of training dataset, CNN component and pre-train setting are demonstrated on both the test and evaluation datasets. It is shown that using the pre-trained CNN components can significantly increase system performance, where the pre-train curves have better SED results on both the test and evaluation datasets. It implies that the knowledge learned from a natural image dataset like 'ImageNet' can be somehow transferred to learn the audio time-frequency features. As more unlabeled in-domain data are added to the training dataset as shown in Table 1, the SED performance generally increases. However, the SED F1 curves also show some fluctuations at the same time, which is due to the noisy label effect of using unlabeled data since more unlabeled data is considered more label noise is added to training where there should be a trade-off. Moreover, the Xception-based SED systems show slightly better performance than the ResNet-based systems. As shown in Table 2, the SED results of fusion systems that combine the best ResNet-based system and Xception-based system are demonstrated for each class. It is found that performance of the fusion system with 160 time-frames is slightly better than that of the system with 120 time-frames. Finally, the proposed system achieves a F1-value as 21.0% which is significantly better than the baseline system which obtains F1 as 10.8%.

**Table 2.** The class-wise F1-measures of fusion SED systems on the evaluation dataset (%)

|  | 160 time-frames | | | 120 time-frames | | |
| --- | --- | --- | --- | --- | --- | --- |
| Class | ResNet | Xception | Fusion | ResNet | Xception | Fusion |
| Alarm/bell | 21.9 | 27.4 | 25.2 | 14.1 | 13.9 | 16.7 |
| Blender | 32.4 | 21.3 | 29.1 | 28.7 | 31.3 | 33.2 |
| Cat | 1.6 | 3.8 | 2.6 | 2.1 | 2.9 | 3.5 |
| Dishes | 1.3 | 3.6 | 2.9 | 1.1 | 1.7 | 1.6 |
| Dog | 3.8 | 6.2 | 7.4 | 3.6 | 4.3 | 4.7 |
| Electric shaver | 28.6 | 31.1 | 35.4 | 37.8 | 38.1 | 42.5 |
| Frying | 5.6 | 7.6 | 8.1 | 9.5 | 11.1 | 10.5 |
| Running water | 21.9 | 19.8 | 23.7 | 19.4 | 18.2 | 20.7 |
| Speech | 28.4 | 35.7 | 33.7 | 24.4 | 27.3 | 30.5 |
| Vacuum cleaner | 37.2 | 38.3 | 42.3 | 43.0 | 41.4 | 42.9 |
| Average | **18.3** | **19.5** | **21.0** | **18.4** | **19.0** | **20.7** |

## 5. CONCLUSIONS

In this work, we have investigated the use of a scalable CRNN-based system integrated with well-developed CNN architectures for SED in a weakly supervised manner. A state-of-the-art audio tagging system developed by us is utilized to make best use of large-scale unlabeled in-domain data. It is found that the SED performance can be significantly increased by using pre-trained CNN weights on 'ImageNet'. In addition, the results show that using more unlabeled in-domain data in training will generally improve the SED results further but requiring a trade-off when the confidence-level of the predicted weak labels is very low. The noisy label problem of unlabeled data with predicted labels can also be reduced by using the ensemble strategy. Based on the evaluation results, the proposed system finally achieves a F1-value as 21.0% which is significantly better than the baseline system.